\documentstyle[aps,preprint]{revtex}
\input epsf

\title{A 2-D asymmetric exclusion model for granular flows}

\author{Christophe Josserand\\
{\small\it The James Franck Institute, The University of Chicago, 
5640 South Ellis Avenue,\\
 Chicago, Illinois 60637, USA}
}

\author{\parbox{430pt}{\vglue 0.3 cm \small  A 2-D version of
the asymmetric exclusion model for granular sheared flows
is presented. The velocity profile exhibits
two qualitatively different behaviors, dependent on control parameters.
For low friction, the velocity profile follows an exponential decay while
for large friction the profile is more accurately represented by a Gaussian 
law. The phase transition occurring between these two behavior is
identified by the appearance of correlations in the cluster size 
distribution. Finally, a mean--field theory gives qualitative 
and quantitative good agreement with the numerical results. 
}}

\begin{document}

\maketitle

\section{Introduction}  
Among the numerous problems dealing with granular materials, one of the
most challenging is granular sheared flow\cite{JNB,raj,leo}. Thus,
although granular materials might exhibit solid, liquid or gas-like 
properties\cite{JNB}, the granular flows cannot be described simply.
Granular sheared flow arises in many different
contexts such as pipe flow, pyroclastic flows\cite{straub}, or even 
traffic jams\cite{eli}. 
Experiments performed in 2-D and 3-D geometries show surprising velocity 
profiles. First, the flow occurs only in a sheared region whose
width is typically on the order of ten particle sizes (the
shear zone). Also, velocity profiles appear
to behave differently in 2 and 3 spatial dimensions. The velocity
decreases exponentially with the distance to the wall in 2-D with
a small Gaussian correction\cite{behr}, while in 3-D the profile is
almost purely Gaussian\cite{mri}. The goal of this paper is to 
present a two dimensional ``toy model'' where the velocity profile 
evolves correspondingly from an exponential like form to an
almost Gaussian form. The model consists of vertically coupled layers.
Each layer follows the well known asymmetric exclusion (ASEP) model. 
In one dimension, the ASEP model has been widely studied and
under certain conditions, exact solutions have been found using
the infinite dimension matrix method\cite{derri}. However, to our
knowledge, very little is known concerning 2-D
ASEP model. Our numerical simulations will in fact show a cross-over 
between an exponential velocity 
profile and a Gaussian velocity profiles when the control parameter 
crosses the value $1/2$. This could be interpreted as a phase transition
in infinite size systems, as the study of the clusters size distribution 
will indicate. Finally a mean field approach will be developed for the
low values of the control parameter.

\section{The model}
\begin{figure}[h]
\centerline{ \epsfxsize=16truecm \epsfbox{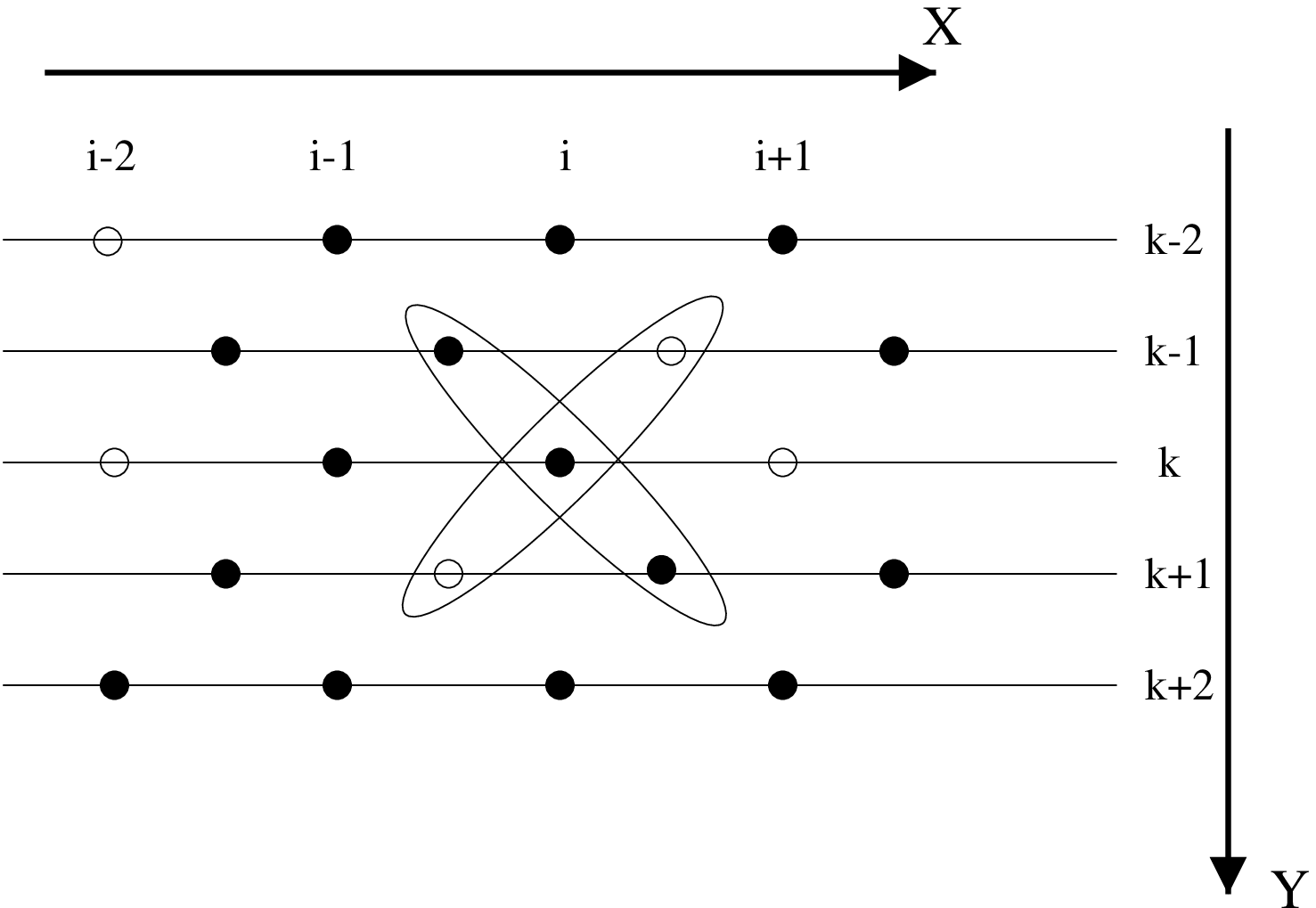} }
\caption{\protect\small 2-D lattice with sites (circle) at time $t_n$: the 
white circles represent the empty sites or holes, while the black
circles represent particles. The motion of the particle located at 
the position $(i,k)$ is determined as indicated by the motion of its
four closest neighbors on rows $L_{k+1}$ and $L_{k-1}$.
\label{schema}}
\end{figure}

The 1-D ASEP model with periodic boundary conditions describes a  
one-dimensional lattice of N sites, where each site $i$ ($1 \le i \le N$) 
is either occupied by a particle or empty.
During time interval $dt$, each particle has a 
probability $dt$ of jumping to the adjacent site to its right if that 
site is empty. Our 2-D has a 
simplified dynamics since the time is now discretized 
($t_n=n$).
Each layer is a one dimensional lattice of $N$ sites, with periodic
boundary conditions. The layers are labeled $L_k$ ($k
\ge 0 $). Each site is either occupied or empty and
 the density along each layer is set to be 
constant ($=\rho$). A shift of one half of the grid spacing is
applied between consecutive layers, so
that for $k$ even $i$ is an integer and for $k$ odd, $i$ is an half integer
(see figure \ref{schema}). The 2-D lattice is composed of an infinite
number of layers occupying the half space $y \ge 0$. At time $t_n$, 
each particle has a probability $P_{i,k}(t_n)$ of hoping to the right.
The quantity $P_{i,k}(t_n)$ is determined by the dynamics of the four 
nearest neighbors of the site $(i,k)$: two in the row $L_{k-1}$ at
time $t_n$ and two in the row $L_{k+1}$ at time $t_{n-1}$. 
$P_{i,k}(t_n)$ is simply proportional to the
number of these neighbors that are moving with a proportionality 
coefficient $\alpha$. In addition, the exclusion 
principle imposes that the particle cannot jump to an occupied
site. However, if the particle on site 
 $(i+1,k)$ is jumping to its right at time $t_n$ then the
 particle on site $(i,k)$ is allowed to move.
If $Q_{i,k}(t)$ is the characteristic function of the motion for the 
site $(i,k)$ at time $t$ ($Q_{i,k}(t)=1$ if there is a particle
at time $t$ that is jumping from site $(i,k)$, and is zero otherwise)
the equation for $P_{i,k}(t_n)$ reads:

\begin{equation}
 P_{i,k}(t_n)= \alpha (Q_{i-1/2,k-1}(t_n)+Q_{i+1/2,k-1}(t_n)+ 
Q_{i-1/2,k+1}(t_{n-1})+Q_{i+1/2,k+1}(t_{n-1}))
\label{defi}
\end{equation}

For consistency, if the right hand side of the formula (\ref{defi}) 
is bigger than $1$ then we define $P_{i,k}(t_n)=1$. Eventually, our 
boundary condition will represent a moving wall situated at
$k=-1$: all the sites of $L_{-1}$ are filled and are moving at each
time step. No boundary condition is required for $k=\infty$.
The velocity $V_k(t_n)$ is defined as the probability that a particle 
located on row $L_k$ moves at time $t_n$. Then, the mean value of 
the velocity on row $L_k$ is denoted $V(k)$ and defines the velocity profile.

The only control parameter of the dynamics is $\alpha$.
 Experimental observations suggest that $\rho$ should be
taken close to $1$. It appears from the simulations that the 
dependence on $\rho$ is trivial, and we consider results for
$\rho=0.9$. On the other hand,
$\alpha$ reveals how much a moving particle pushes on its neighbors.
 Therefore, it has to be a (non trivial {\it a priori}) 
function of the material properties such as the friction, the 
shape of the grains, their roughness, etc...

The model does not allow exchange of particles from one row to another
(along the $y$ direction). This strong constraint is in 
opposition with the experimental observations, where the density 
profile seems to reach a steady state where the exchanges between rows 
just balance. We assume in fact in this model that
this interchange of particles is not relevant compared to the friction effects.
We also checked that by imposing a reasonable density profile from 
the moving boundary to the bulk, the qualitative results of the model are not
affected.

\section{Numerical results.}

\begin{figure}[h]
\centerline{ \epsfxsize=16truecm \epsfbox{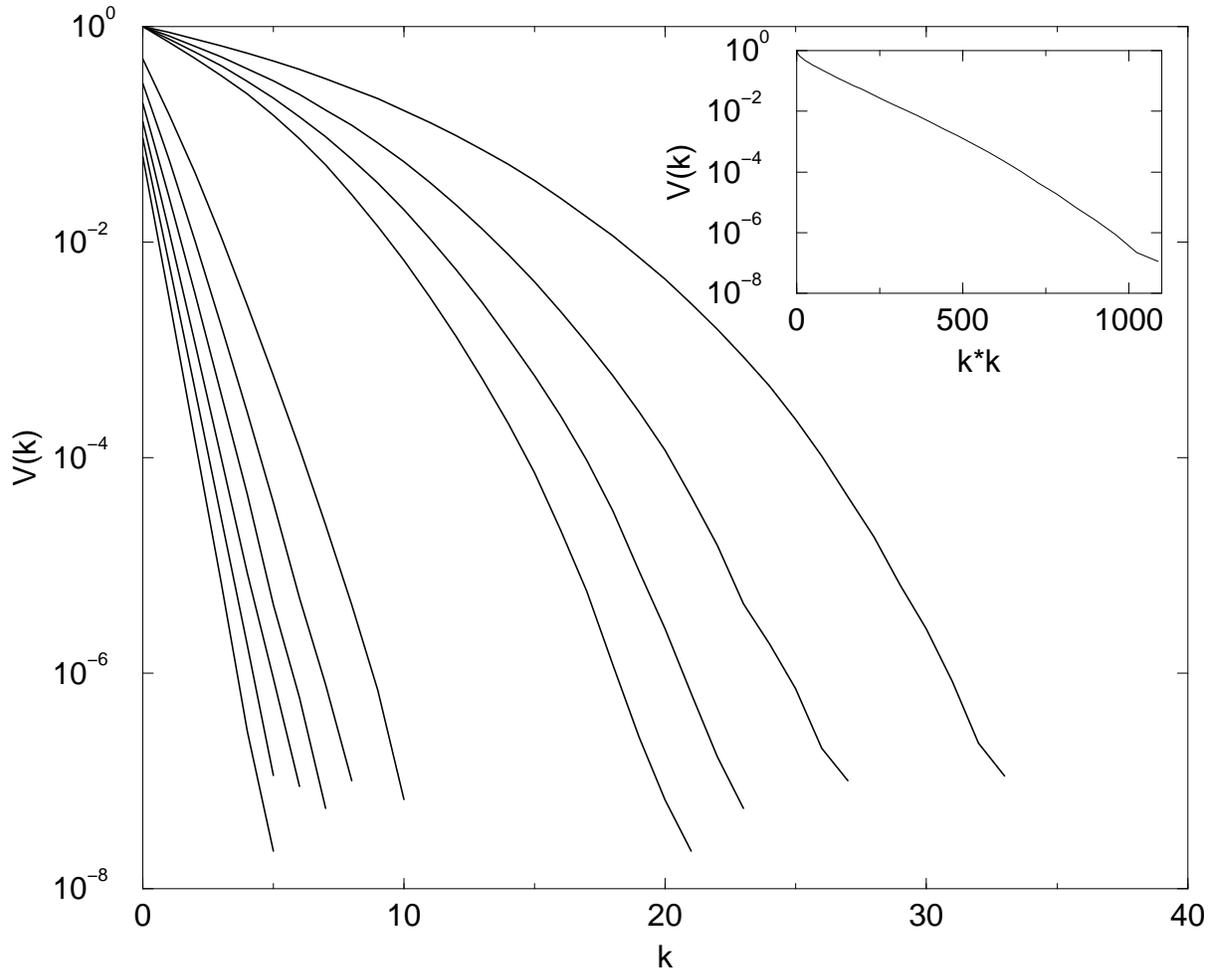} }
\caption{\protect\small Velocity profile for a density $\rho=0.9$ for
different values of the control parameter $\alpha$. $\alpha$ increases
from $0.2$ to $0.65$ in increments of $0.05$ as the curves are plotted
from the left to the right. For $\alpha$ smaller than one half the 
profile is almost exponential and it is far from exponential for 
$\alpha \ge 0.5$. The insert shows the velocity profile for
$\alpha=0.65$ as a function of $k^2$. It shows 
that the profile is better approximated by a Gaussian law, particularly for 
the low values of $k$. The number of sites per row is $20000$ and the
average is computed over $4000$ time iterations for $5$ different 
initial conditions for each $\alpha$.
\label{profil}}
\end{figure}

Figure (\ref{profil}) shows different velocity profiles for $\rho=0.9$
and $\alpha$ increasing from $0.2$ to $0.65$. For $\alpha \ge 1/2$,
an abrupt change in the velocity profile is observed. 
In fact, we can expand the logarithm of the velocity in a Taylor serie:
\begin{equation}
 ln(V(k))= a +b \cdot k+ c \cdot k^2 +d \cdot k^3...
\label{varfit}
\end{equation}
We remark that the three first terms on the right hand side
of (\ref{varfit}) give 
a qualitatively good approximation of the velocity profiles. It 
correspond to a Gaussian fit of the profile. However,
the ratio $b/c$ indicates the typical width for which the quadratic term 
becomes of the same order as the linear one. For $\alpha <0.5$ 
this ratio is of order of hundreds, {\it i.e.} much larger than the shear
width and the dynamics can be considered almost purely exponential. 
 On the other hand, the ratio $b/c$ becomes of the order of $1$ when 
$\alpha$ crosses the critical value $\alpha_c=0.5$, so that one can 
approximate the velocity profiles for $\alpha>\alpha_c$ with a 
Gaussian centered near the row $k=0$. Such property appears clearly 
on the insert of figure (\ref{profil}), where the velocity profile 
for $\alpha=0.65$ as a function of $k^2$ is shown.
 Notice that for small $k$ ($k^2 <500$), the logarithm 
of the velocity is almost linear in $k^2$, so that one can consider 
that the velocity profile is Gaussian at least near the wall.

The instantaneous velocity $V_k(t)$ shows also 
different behaviors wether $\alpha$ is larger or smaller than
$\alpha_c$. Thus, for $\alpha \ge 0.5$, transitory dynamics 
occur for small time ($t_n < 100$) until the velocity reaches a
stationary behavior. Such transitory states cannot be seen for
$\alpha$ smaller than $0.5$.

In order to investigate more carefully the transition occurring 
at $\alpha=\alpha_c$, as well as the velocity profiles,
a simplified version of the model is introduced. It exhibits 
the same properties as
the model explained above, and can be more easily studied 
analytically. It consists of neglecting the effect of the row 
$L_{k+1}$ on the dynamics in row $L_k$. The model breaks 
the symmetry along the $y$ direction. However, experimentally, 
the sheared flows exhibit a spontaneous symmetry breaking as well.
A transition from exponential to Gaussian velocity profile 
occurs in the same way for this simplified model at
$\alpha=\alpha_c$. Notice that $\alpha_c$ corresponds to the value 
of $\alpha$ at which the probability becomes $1$ to move if
the two neighbors above are moving (the exclusion condition still
holds).

\begin{figure}
\centerline{  \epsfxsize=16truecm \epsfbox{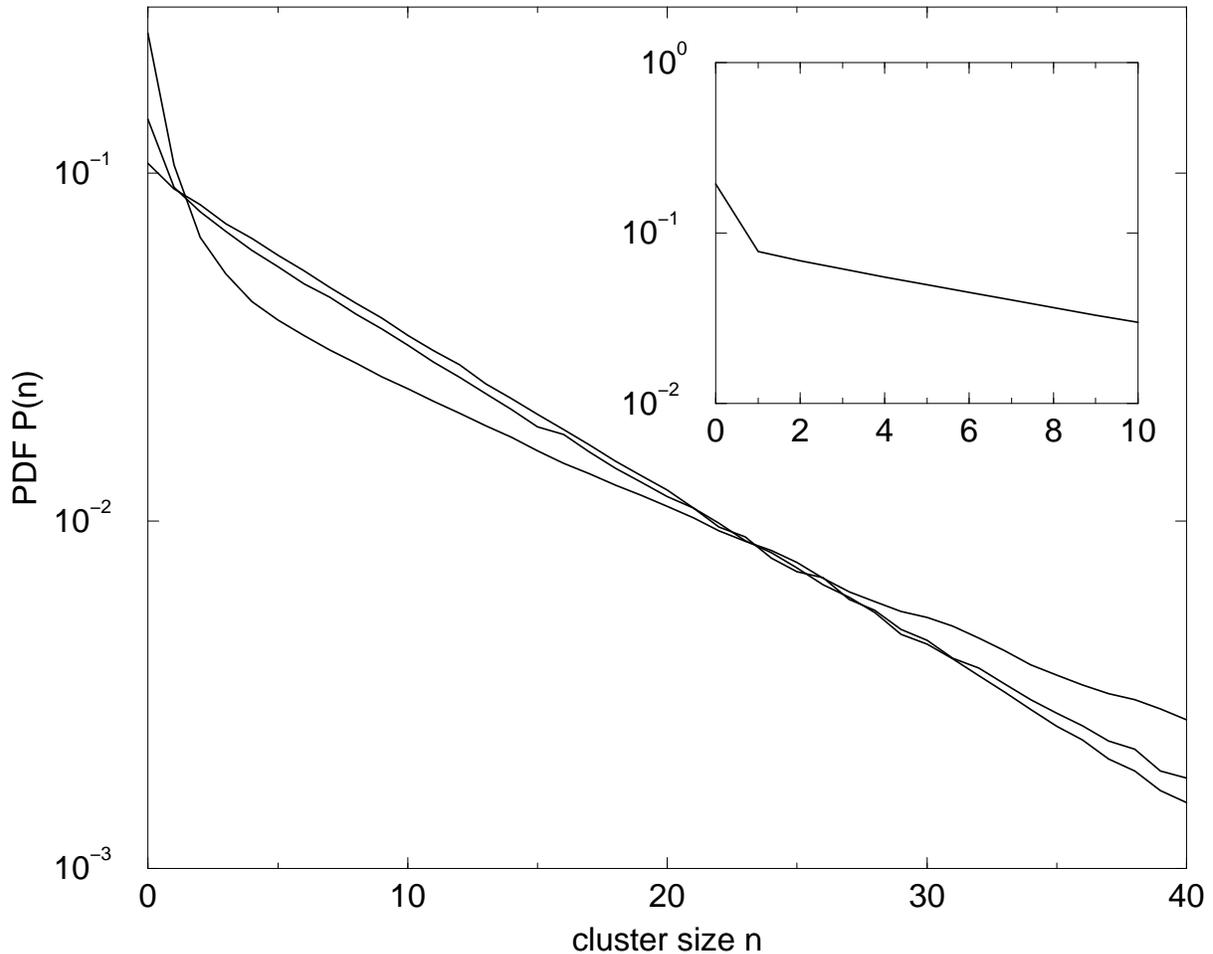}  }
\caption{\protect\small The probability distribution function of
the cluster size for different values of $\alpha$ and $k$, and for
$\rho=0.9$. The 
straight line corresponds to $P_0(n)$ and is $P(n)$ for
$\alpha=0.45$ and $k=1$. The two others PDF shown are for value
of $\alpha$ larger than $\alpha_c$. The one that is maximum for $n=0$ 
is for $\alpha=0.5$
and $k=1$ while the nest highest is for $\alpha=0.6$ and $k=4$. Notice 
that they show Poisson-like tails at large $n$, with different 
slopes. Insert: for $\alpha=0.49$ and $k=1$, the PDF differs from 
$P_0(n)$ only at $n=0$.
\label{pdf}}
\end{figure}

We define $P(n)$ as the probability that, given a hole, the next 
hole on its right is located after $n$ filled sites. 
$P(n)$ gives the probability distribution function (PDF) for the size of the
clusters (if two consecutive sites are empty, it is considered as
a cluster of size $0$). {\it A priori}, the function $P(n)$ depends 
on $\rho$, $\alpha$ and the row number $k$. However, if we
consider that the particles are placed randomly on the sites with 
a density $\rho$ (as do the initial conditions), one obtains the Poisson 
distribution $ P_0(n)=(1-\rho) \rho^n $. As shown on figure (\ref{pdf}), 
for $\alpha<\alpha_c$, $P(n)$ corresponds almost exactly to $P_0(n)$ 
for any $k$ and $\alpha$. Therefore, at each time step, the 
particles are distributed on the sites as if they were randomly 
placed with density $\rho$. On the other hand, for $\alpha \ge \alpha_c$, 
the function $P(n)$ differs from $P_0(n)$ in that the small 
size clusters are more frequent, while large clusters follow a 
Poisson-like law. In this case, $P(n)$ does not vary as $\alpha$
increases for a given row, but changes as the row number increases: 
the bigger $k$, the closer $P(n)$ approaches $P_0(n)$.

\begin{figure}
\centerline{  \epsfxsize=16truecm \epsfbox{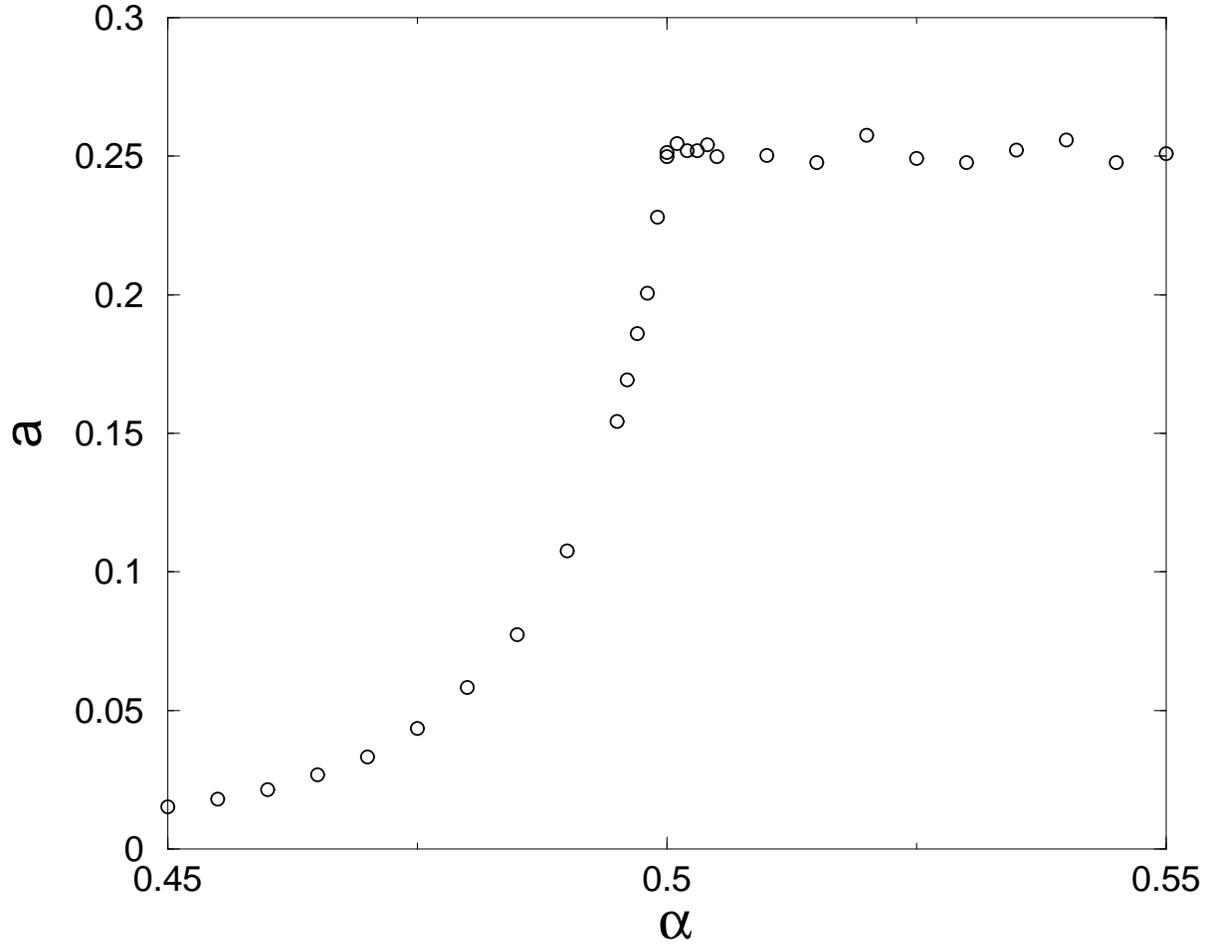}  }
\caption{\protect\small The density of holes $a$ as a
function of $\alpha$ near the phase transition
($\alpha=\alpha_c=0.5$). $a$ is approximately constant for 
$\alpha >\alpha_c$ and exhibits a critical exponent of $2/3$ for 
$\alpha \le\alpha_c$.
\label{dista}}
\end{figure}

When $\alpha$ approaches $ \alpha_c$ the PDF differs slightly
from $P_0(n)$, so for $k=1$ we can 
expand $P(n)$ in the form (see insert of figure \ref{pdf}):
$$ P(n)=a \delta_{n,0}+(1-a) \rho_{eff}^n (1-\rho_{eff}), $$
with $\rho_{eff}=\rho/(1-a(1-\rho))$ (mean density has
to be $\rho$). Figure (\ref{dista}) shows the evolution of $a$ as
a function of $\alpha$ for $\rho=0.9$. For $\alpha>\alpha_c$, $a$ is
a constant since $a=a_c=0.25 \pm 0.01$ which corresponds to
$\rho_{eff} \sim 0.92$. However, for $\alpha \le \alpha_c$ we 
observe that for $a\sim a_c$:

$$ a_c-a \propto (\alpha_c-\alpha)^{\nu} $$
and we found $\nu \sim 2/3$. Therefore, we identify the 
behavior of the system when 
$\alpha$ reaches $\alpha_c$ as a phase transition. Notice that 
when $\alpha$ increases above $\alpha_c$, the PDF shows 
a more complex behavior since not only 
the value of $P(0)$ is disturbed, but also a larger range of small
clusters size $n$ (see figure \ref{pdf}).

\section{Mean field theory}

As the correlations can be neglected for $\alpha < \alpha_c$, a 
mean field approach is appliable. Defining
$P^{(k)}(n)$ as the probability that $n$ successive 
particles are moving on the row $L_k$ (knowing that
an empty site is before such a cluster),
 $P^{(0)}(n)$ can be computed exactly: $ P^{(0)}(n)= 
(2\alpha\rho)^n (1-2\alpha \rho) $
and by noticing that $V(0)=\frac{(1-\rho)}{\rho} \sum nP^{(0)}(n) $, 
we obtain:
$$ V(0)=\frac{2\alpha (1-\rho)}{1-2\alpha \rho} $$
It is remarkable that the numerical simulations and this 
mean--field solution agree within an error of
less than one percent. Also, one can write the constitutive relation
between $P^{(0)}$ and $P^{(1)}$:

\begin{equation}
P^{(1)}(n)=\sum_{l=n-1}^\infty W(l \rightarrow n) P^{(0)}(l)
\end{equation}
where $W(l \rightarrow n)$ is the probability that if a
cluster of size $l$ at row $k=0$ is moving, then a cluster
 of size $n$ at row $k=1$ is moving. It follows that:
$$ W(l \rightarrow n)= (1-\rho)(2\alpha\rho)^n \left( (l-n+1)-(l-n)
2\alpha \rho \right) \quad {\rm for} \quad l>n $$
$$ W(n \rightarrow n)=(1-\rho)\alpha \rho (2\alpha\rho)^{n-1}(2-\alpha\rho);
 W(n-1 \rightarrow n)=(1-\rho)(\alpha\rho)^2 (2\alpha\rho)^{n-2}$$
and we obtain for $V(1)$:

$$ V(1)=\frac{1-\rho}{\rho} \sum_{n=1}^\infty nP^{(1)}(n)=
\frac{2\alpha^2 \rho (1-\rho)^2(2+\alpha\rho(1-2\alpha\rho)^2)}
{(1-2\alpha\rho)^2(1+2\alpha\rho)}. $$

Again, this mean--field solution is in good agreement with the
numerical results, although not as accurate than for $V(0)$.
Unfortunately, the next order step, for obtaining the
analytical solutions of $P^{(1)}$ and then $V(2)$ are 
much more complicate. However, one can consider that for
the exponential behavior, the knowledge of $V(0)$ and $V(1)$ 
is sufficient.
Then, we can compare the numerical exponential profile with the
exponential law predicted by the mean--field.
We define $r(\alpha,\rho)$ such that for $\alpha < \alpha_c$,
the velocity profile obtained numerically is fit by:
$ V(k)=V(0) (r(\alpha,\rho))^k $

\begin{figure}[h]
\centerline{ \epsfxsize=16truecm \epsfbox{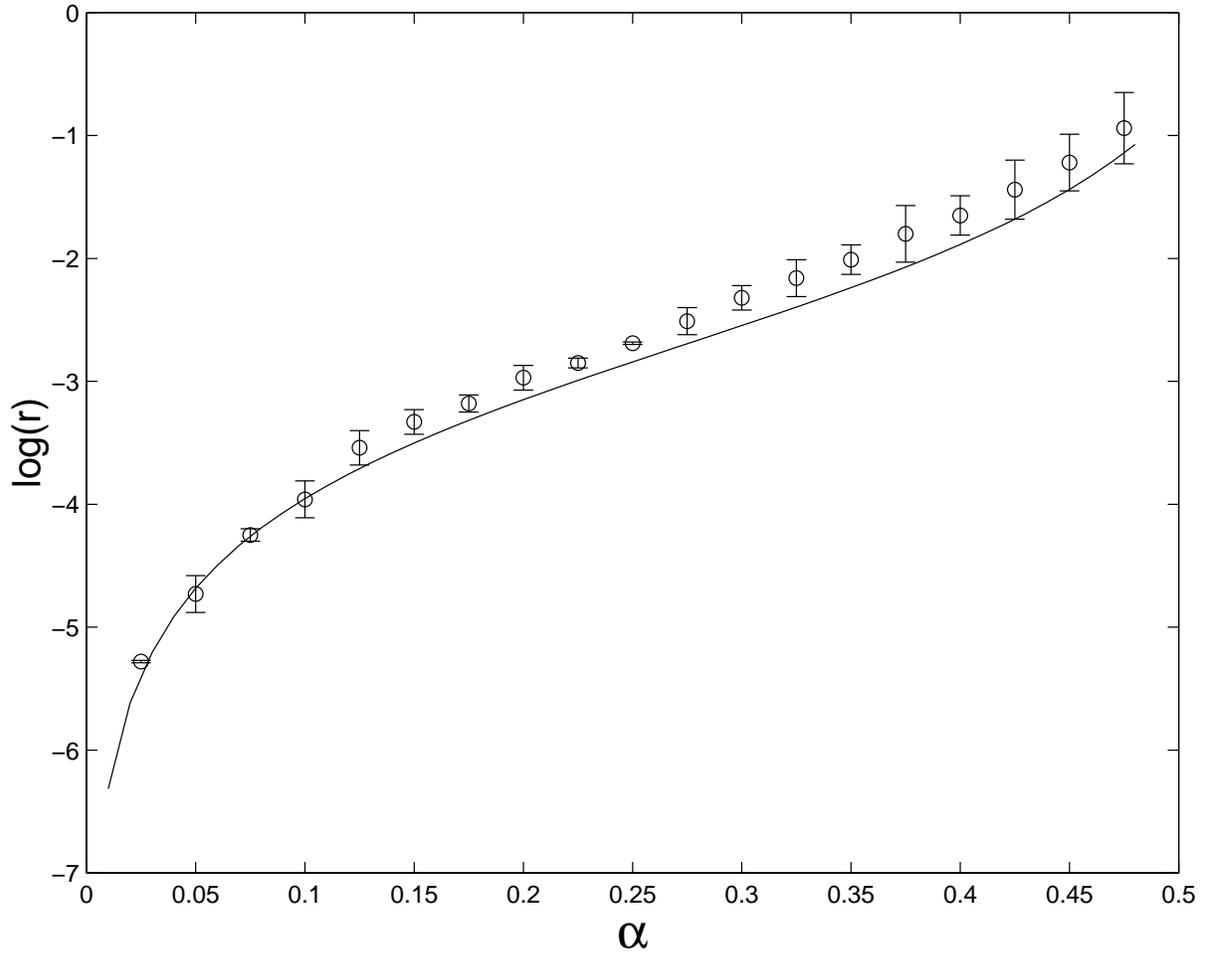} }
\caption{\protect\small  $log(r(\rho,\alpha))$ obtained by numerical 
simulation (circles) for $\rho=0.9$ compared to the mean--field
approximation $log(V(1)/V(0))$ (line). $log(r)$ is computed
by a Gaussian fit of the numerical results. The error bars have been
evaluated by comparing it with an exponential fit.
\label{pente}}
\end{figure}

Figure (\ref{pente}) shows $log(r)$ as function of $\alpha$ for
$\rho=0.9$, compared with the mean--field $r_{mf}$ solution taken as:

$$ r_{mf}=\frac{V(1)}{V(0)}=\frac{\alpha \rho (1-\rho)(2+\alpha\rho
(1-2\alpha\rho)^2)}{1-(2\alpha\rho)^2} $$

Thus, the mean--field approximation, deducing the exponential law from the 
ratio between the two first velocities, gives a quantitative good
approximation of the exponential decay for $\alpha<\alpha_c$.

But, for $\alpha \ge \alpha_c$, the mean--field approach fails and we
are not able to show analytical results. Although we were able to 
quantify the coorelations for $\alpha \sim \alpha_c$, numerical 
simulations show correlations along the $y$ direction as well.
However, we believe
that the phase transition exhibited by the model might have 
interesting features in granular flows. Particularly, it might be
interesting to perform an experiment where the friction coefficient 
of the granular materials would change. Also, the model shows a 
phase transition between exponential and Gaussain behavior as 
$\alpha$ increases, while experimentally, the different behaviors
occur between 2 and 3 spatial dimensions. In fact, one can imagine,
for example, that the 3-D experiment can be linked to this 2-D
ASEP model with an effective friction coefficient $\alpha_{3D}$.
Then, if one consider that in 3-D each particle has more 
neighbors that can push it, it is plausible to assume that $\alpha_{3D}>
\alpha_{2D}$ and possibly $\alpha_{3D} > \alpha_c \alpha_{2D}$ under
certain conditions.
Also, we would like to point out other applications of this model
such as non-newtonian flows or molecular frictions.

Finally, notice that he model is based mainly on probabilistic
properties of granular flows. Other stochastic approaches have 
already been proposed in this context\cite{pouliq,dejos}, although 
in granular materials there is no justification such as thermal 
fluctuations for stochastic processes. However,
the shape of the grains, and therefore the contact network in granular 
materials can be considered as random variables (this has been argued 
for the chain forces in static bead pile\cite{sue}). Additionally, 
the motion of the particles in shear flows changes 
the configuration of the contact network of the system.

It is my pleasure to thank Eli Ben-Naim, Dan Mueth, Georges Debregeas
and Leo Kadanoff for their advice and their interest on this work.
I also acknowledge the ONR (grant: N00014-96-1-0127), the
MRSEC with the National Science Foundation DMR grant: 9400379 and
the ASCI Flash Center at the University of Chicago under DOE 
contract B341495 for their support.

\newpage

\newpage

\newpage

\newpage

\newpage

\end{document}